# Moiré-modulated band gap and van Hove singularities in twisted bilayer germanene

Pantelis Bampoulis[1,&], Carolien Castenmiller[1], Dennis J. Klaassen[1], , Jelle v. Mil[1], Paul L. de Boeij[1], Motohiko Ezawa[2] and Harold J.W. Zandvliet[1]

[1] Physics of Interfaces and Nanomaterials, MESA+ Institute for Nanotechnology, University of Twente, P.O. Box 217, 7500AE Enschede, The Netherlands
[2] Department of Applied Physics, University of Tokyo, Hongo, 113-8656 Tokyo, Japan

Twisting bilayers of two-dimensional topological insulators has the potential to create unique quantum states of matter. Here, we successfully synthesized a twisted bilayer of germanene on Ge$_2$Pt(101) with a 21.8° degrees twist angle, corresponding to a commensurate ($\sqrt{7}\times\sqrt{7}$) structure. Using scanning tunneling microscopy and spectroscopy, we unraveled the structural and electronic properties of this configuration, revealing a moiré-modulated band gap and a well-defined edge state. This band gap opens at AB/BA stacked sites and closes at AA stacked sites, a phenomenon attributed to the electric field induced by the scanning tunneling microscopy tip. Our study further revealed two van Hove singularities at -0.8 eV and +1.04 eV, resulting in a Fermi velocity of $(8 \pm 1) \times 10^5$ m/s. Our tight-binding results uncover a unique quantum state, where the topological properties could be regulated through an electric field, potentially triggering two topological phase transitions.

&) Corresponding author: p.bampoulis@utwente.nl

**Introduction**

Moiré superlattices, formed by vertically stacking two-dimensional (2D) materials, offer a platform where unique quantum phenomena can materialize, attributed to unexpected alterations in electronic band structure and topology [1-9]. The properties of these superlattices are not an aggregate of their constituent layers, but a complex result of factors including interlayer coupling, lattice reconstruction, and twist angle. The strength of the interlayer coupling is vital, facilitating electron hopping and charge redistribution between layers, thus fostering their unique properties. In such stacked bilayers the twist angle introduces an additional degree of freedom. If the top layer is twisted with respect to the bottom layer, a moiré superstructure emerges. The periodicity of this moiré structure is given by $\lambda = a/\left(2sin\left(\frac{\vartheta}{2}\right)\right)$, where $a$ is the lattice constant of the 2D material and $\vartheta$ the twist angle between the two layers. In graphene, the twisting leads to a shift of the two Dirac points of the two layers in reciprocal space, resulting in a crossing of the Dirac cones. Owing to this crossing, novel electronic states emerge in the vicinity of the Fermi level. These electronic states result in van Hove singularities in the density of states [8,10]. These van Hove singularities, which have an appreciable density of states, can be brought arbitrarily close to the Fermi level by tuning the twist angle or by gating, leading to electronic instabilities and correlated electronic phases, such as superconductivity, Mott metal-insulator transition, tuneable magnetism, Wigner crystallization etc. [9,11,12].

Twistronics, traditionally focused on graphene and graphene-based 2D structures [1], are now expanding to incorporate non-graphene 2D materials like silicene and bismuthene [13-16]. This shift promises potential discoveries of unique topological phases and kagome flat band systems. Particularly, when considering 2D topological insulators, the interplay of topology and electron correlations offer fertile grounds for the emergence of exotic phases [17-18]. In the same context, germanene – the germanium analogue of graphene with a buckled honeycomb lattice [19] – offers promising prospects with its Dirac fermions [20,21] and strong spin-orbit coupling [22,23]. In our recent work, we demonstrated that germanene grown on $Ge_2Pt$ [20] is a 2D topological insulator, exhibiting the quantum spin Hall effect [21]. Notably, germanene hosts a gapped interior with spin-polarized topologically protected edge states at its boundaries, and undergoes a topological phase transition when a perpendicular electric field is applied.

Inspired by these findings and the potential of twisting a 2D topological insulator, we have successfully grown twisted bilayers of germanene using the spinodal decomposition of a GePt eutectic alloy. The twisted bilayers have a twist angle of 21.8º, corresponding to a commensurate $(\sqrt{7} \times \sqrt{7})$ structure. Using scanning tunneling microscopy and spectroscopy (STM, STS) at 77 K, we have probed the structural and electronic properties of these twisted bilayers. Our findings reveal a moiré-modulated band gap, a metallic edge state, two van Hove singularities, and a Fermi velocity of (8±1)×10$^5$ m/s. Remarkably, our tight binding calculations indicate that an applied electric field could trigger two topological phase transitions.

**Results and Discussion**

Numerous research groups have reported the growth of germanene by depositing Ge atoms on metal substrates, such as Pt(111) [24], Au(111) [25], and Al(111) [26]. Initially, the assumption was that these produced structures were pure layers of germanene. However, more recent investigations have suggested that these structures are in fact alloys, comprised of Ge atoms mixed with the atoms of the respective metal substrate [27-29]. Our methodology, which relies on spinodal decomposition of eutectic Ge$_{0.78}$Pt$_{0.22}$ droplets into Ge$_2$Pt and pure Ge, stands apart from this prevalent alloying issue. This method circumvents the problem of alloying as it utilizes all available Pt atoms already for the formation of the Ge$_2$Pt substrate. The surplus Ge, in this case, is expelled onto the surface of the Ge$_2$Pt crystals, ensuring the formation of pure germanene layers. The growth of germanene on Ge$_2$Pt is based on the methodology we established in previous research [20,21]. Initially, Pt is deposited onto a Ge(110) surface in ultra-high vacuum. Subsequently, the system is heated just above 1047 K to create Ge$_{0.78}$Pt$_{0.22}$ eutectic droplets. The eutectic droplet formation process is extensively discussed in refs. [30-32]. Upon returning to room temperature, the spinodal decomposition of these droplets occurs, leading to a crystalline Ge$_2$Pt phase and a distinct pure Ge phase [20,30]. This process results in the formation of germanene layers on Ge$_2$Pt(101), which are separated by a buffer layer, as elaborated in ref. [21]. The germanene layers on top of the buffer layer on Ge$_2$Pt exhibit a buckled honeycomb structure. This buckling is in prefect agreement with theoretical calculations [19]. We have recently demonstrated that germanene on a buffer layer/Ge$_2$Pt substrate is a two-dimensional topological insulator that undergoes a topological phase transition upon the application of an electric field [21]. At the criticial field, the gap closes and germanene is characterized by low-energy linear bands [21, 33].

Figure 1(a) shows a large-scale image of the germanene/Ge$_2$Pt(101) surface. The STM image displays two distinct levels with a separation of 0.28 nm, matching the monoatomic step height of germanene, evident from the line profile in the inset of Figure 1(a). Figure 1(b) presents a schematic representation of the germanene/Ge$_2$Pt system, indicating that the first germanium layer on Ge$_2$Pt(101) is electronically dead. This germanene buffer layer decouples the other germanene layers from the underlying substrate, see also ref. [21]. An atomic-scale STM image and its corresponding Fast Fourier Transform (FFT) image of the buffer layer on Ge$_2$Pt showcases a complex unit cell, see the left panels of Figure 1(c). This complexity is understood by superimposing the honeycomb lattice of germanene with the rhombic lattice of Ge$_2$Pt's upper Pt layer, see ref. [21] for more details. The difference between these two lattices results in a one-dimensional moiré pattern, identifiable in the STM image as well as in the FFT image as the yellow marked spots in Figure 1(c) [34]. The subsequent germanene layer on the buffer (referred to as the first layer, 1L) demonstrates topological properties [21]. Structurally, this layer's buckled honeycomb lattice is clearly visible in the right panels of Figure 1(c). In addition, the buffer/Ge$_2$Pt moiré pattern disappears and the Ge$_2$Pt spots are barely visible. Germanene on the buffer layer has a lattice constant of ~4.2 – 4.3 Å and a buckling of ~0.4 Å, as extracted from atomic resolution STM images.

The predominant structure of the few-layer germanene atop the buffer layer on Ge$_2$Pt (101) typically adopts an AA stacking pattern, as shown in the right panel of Figure 1(c). Density functional theory calculations have shown that AA stacking is more favorable than AB stacking [19]. Approximately 10% of few-layer germanene/buffer/Ge$_2$Pt areas exhibit a distinctive and large superstructure. This is illustrated in the large-scale STM image of Figure 1(d) and the detailed close-up in Figure 1(e). The superstructure's unit cell, as depicted in Figure 1(e), has a dimension of ~1.1 nm, which is significantly larger than the lattice constant of monolayer germanene (right panel of Figure 1(c)). In order to unravel the exact details of the superstructure we performed an FFT of the STM image in panel (e). The FFT image, shown in Figure 1(f), reveals many spots that can all be explained by a twisted germanene bilayer. The yellow hexagon represents the reciprocal lattice of the (1x1) germanene unit cell. The size of the white hexagon is smaller by a factor $\sqrt{7}$ and refers to the reciprocal lattice of the ($\sqrt{7} \times \sqrt{7}$) moiré unit cell of the twisted bilayer germanene. Replicas of the moiré unit cell are visible around the (1x1) spots of the germanene unit cell (outlined by dashed white lines). The rotation angle between the moiré unit cell and the honeycomb lattice of the top germanene layer, $\theta$, is estimated to be ~$17 \pm 1^o$. The twist angle between the two germanene layers, $\vartheta$, can be determined using the relation $\vartheta = 30^o - \theta/2$ and amounts ~$21.5 \pm 0.5^o$, i.e. consistent with 21.8$^o$, which is one of the commensurate angles. For specific twist angles, such as 21.8$^o$ and 38.2$^o$, the moiré structure is commensurate with the honeycomb lattice and forms a perfectly periodic structure [35-38]. Twist angles of 21.8$^o$ and 60$^o$-21.8$^o$=38.2$^o$ both have a ($\sqrt{7} \times \sqrt{7}$) moiré unit cell, but they have opposite sublattice-exchange parities. The structure with odd parity exhibits a $C_3$ symmetry, whereas the structure with even parity has a $C_6$ symmetry. Structurally we cannot discriminate between the two. It is worthwhile to mention here that 21.8$^o$ is also a preferred twist angle of twisted bilayer silicene [13,14].

Figure 1(g) presents an atomic resolution STM image, showcasing the unique structure of twisted germanene. Within this image, both the moiré superlattice and the 1x1 lattice of germanene are distinctly observed. In Figure 1(h), we provide a structural model of the twisted bilayer. The model depicts two layers of germanene with a twist angle of 21.8$^o$. This model aligns closely with the actual STM image displayed in Figure 1(g), serving as a guide for identifying different stacking configurations within the superlattice. The model helps to identify and categorize the superlattice sites. AA sites occur at locations where the germanium atoms of the top and bottom layers align directly above each other. Sites labeled as AB have an atom in the top layer in between a honeycomb cell of the bottom layer. Conversely, BA sites have an atom in the bottom layer in between a honeycomb cell of the top layer. Therefore, the AB sites are brighter than the BA sites in an STM image, see Figure 1(g).

The electronic properties of twisted bilayer germanene substantially differ when compared to 1L germanene. Figure 1(i) shows differential conductivity spectra (*dI(V)/dV*) obtained using a lock-in amplifier with a modulation voltage of 20 mV and a frequency of 1.1 kHz. The spectra are shown for twisted bilayer germanene (in red), 1L germanene (in blue), and, for a baseline comparison, on $Ge_2Pt$ (in black). The topological state of 1L germanene (and thus its local density of states) can be influenced by the electric field induced by the STM tip [21]. We have chosen a tip-induced electric field of 1.95 V/nm, which corresponds to a semimetallic state [21], as evident from the V-shape *dI(V)/dV* in Figure 1(i). In stark contrast, the twisted bilayer germanene, under a similar electric field, exhibits a significant gap-like feature in its spectrum, indicating different electronic properties. For context, the $Ge_2Pt$ surface shows a featureless metallic *dI(V)/dV* spectrum.

In order to determine the origin of the band gap of twisted bilayer germanene, we have recorded spatially resolved *dI(V)/dV* spectra along the moiré unit cell. Figure 2(a) compares *dI(V)/dV* curves recorded at AA and AB/BA sites, see inset for the location. The *dI(V)/dV* spectrum of the AB/BA sites reveals a gap of about 0.15 eV. The middle of the gap is located at about 0.07 eV above the Fermi level. In contrast, the AA sites are metallic and exhibit a V-shape density of states, similar to 1L germanene, but the Dirac point is about 0.02 eV above the Fermi level, see Figure 1(i). For the determination of the gap we refer to the supplementary information. In Figure 2(b) we show a *dI(V)/dV* line spectroscopy map taken along the green line shown in the inset of Figure 2(a). It shows that the size of the band gap is modulated by the moiré pattern. The chirality of the Dirac fermions prevents the opening of a band gap [39], thus our *dI(V)/dV* results, have to be explained by another mechanism. Our STM experiments reveal that the moiré-modulated band gap opening is caused by the electric field in the STM tunnel junction. This electric field originates from the difference in work function between the STM tip and the substrate [21,40]. The work function of a Pt/PtIr scanning tunneling microscopy tip is ~ 5.7 eV, whereas the germanene substrate has a work function of ~ 4 eV [40]. For a typical tip-substrate distance of ~0.85 nm we find an electric field of ~2 V/nm [21]. Tight-binding calculations show that the application of an electric field normal to the germanene layers leads to the opening of a band gap in the AB and BA stacked regions, see refs. [41,42]. The size of this bandgap (*Δ*) is given by [43],

$$\Delta = \sqrt{((e^2V^2t_\perp^2)/(t_\perp^2 + e^2V^2))} \tag{1}$$

where $V$ is the applied electrostatic potential across the two germanene layers, $t_\perp$ the interlayer hopping parameter and $e$ the elementary charge. For an interlayer hopping parameter in the range of 0.2 to 0.3 eV and an interlayer bias of 0.2 V to 0.5 V, the bandgap lies in the range of 0.15 to 0.25 eV, in good agreement with the experimentally obtained gap. We note that exact matching of the experimental values with the calculations is difficult due to offsets in estimating the tip-induced electric field [21]. In contrast, and for the same electric field, in AA regions the topological gap closes making germanene a topological semimetal owing to the buckling, in line with ref. [21]. In Figure 2(c), we present *dI(V)/dV* spectra from the edge of a twisted flake, comparing it to the AA and AB/BA regions. The edge is metallic, characterized by an edge state that fills the AB/BA gap. As we discuss in the next section our tight-binding calculations suggest that this state is topological in nature.

Next, we investigate the density of states of twisted bilayer germanene at higher energies. Figure 2(d) shows averaged spectra (each spectrum is the result of averaging a few hundred *dI(V)/dV* spectra on both AA and AB/BA sites) recorded at different locations on the twisted bilayer germanene. The spectra in Figure 2(d) exhibit distinct peaks at approximately -0.8 *eV* and +1.04 *eV*. These two peaks are identified as Van Hove singularities that arise due to the crossing of the Dirac cones of the top and bottom germanene layers [44-47]. In Figure 2(e) we show a schematic diagram of the electronic band structure of twisted bilayer germanene (without spin-orbit coupling). The twist angle results in a shift $\Delta K = \Gamma K_1 - \Gamma K_2 = |\Gamma K_{1,2}|sin(\vartheta/2)$, where $K_{1,2}$ refers to the $K$ points of the top and bottom germanene layer, respectively. The crossing of the two Dirac cones results in two additional states that are symmetrically located on both sides of the Dirac point in energy scale. The energy gap between these two van Hove singularities is referred to as the Van Hove gap, $\Delta E_{VHS}$, and amounts to ~1.84 eV. $\Delta E_{VHS}$ is given by [8,48,49],

$$\Delta E_{VHS} = 2\hbar v_F |K| sin(\vartheta/2) - 2w \qquad (2)$$

where $\hbar$ is the Dirac constant, $v_F$ the Fermi velocity, $\vartheta = 21.8^o$ the twist angle between the two germanene layers, $|K| = 4\pi/3a$ is the length of the $\Gamma K$ vector, $a = 4.2$ Å the lattice constant of germanene and $w$ the hybridization energy [48]. A good estimate of the hybridization energy is $w \approx 0.4 t_\perp \approx 0.12$ eV (assuming an interlayer hopping parameter $t_\perp = 0.3$ eV ). Therefore, the Van Hove gap of 1.84 eV results in a Fermi velocity $v_F = (8 \pm 1) \times 10^5 \, m/s$, which agrees very well with other experimental studies [50,51].

In addition to the two Van Hove singularities in our *dI(V)/dV* spectra, smaller peaks/shoulders appear between the Van Hove singularities and the Dirac point, see

Figure 2(d). In order to scrutinize the low-energy density of states of twisted bilayer germanene and to understand these features, we performed tight-binding calculations. The $(\sqrt{7} \times \sqrt{7})$ moiré unit cell contains 28 germanium atoms, i.e. 14 germanium atoms per germanene layer. In the absence of interlayer interaction we only find two prominent Van Hove singularities owing to the crossing of the Dirac cones of both germanene layers, see supplementary information. When interlayer interaction is considered, the density of states exhibits the two Van Hove singularities (Figure 2(f) and 2(g) for comparing theory and experiments), and a number of additional peaks (peaks in Figure 2(f) in between the two Van Hove singularities), matching the experimental results in Figure 2(g). This is caused by (1) the band folding due to the $(\sqrt{7} \times \sqrt{7})$ superstructure and (2) the band anti-crossing due to the interlayer hopping [52]. The strength of the additional peaks increases with increasing interlayer coupling, see supplementary information. The appearance of broad peaks with several shoulders in the differential conductivity is in qualitative agreement with our tight-binding calculations. Furthermore, our tight-binding results reveal that twisted bilayer germanene can undergo two topological phase transitions as a function of an applied electric field. The key results are summarized in the phase diagram shown in Figure 3.

The quantum matter of state of the AB/BA regions of twisted bilayer germanene depends on the exact values of the applied electric field and interlayer coupling. For a vanishing interlayer coupling the twisted bilayer behaves as two decoupled single layers and therefore exhibits only a single topological phase transition [21], whereas for a non-zero interlayer coupling the twisted bilayer undergoes two topological phase transitions, provided that the interlayer coupling is not too strong, see Figure 3. Without interlayer interaction, the topological number, $Q$, of the germanene bilayer is 2, and hence, there is only one topological phase transition of a two-dimensional topological insulator with *Q=2* to a trivial insulator with *Q=0* [52,53,54]. If interlayer interaction is included, the system becomes more complex. With increasing electric field, twisted bilayer germanene undergoes a series of topological phase transitions. First, the topological number changes from *Q=2* to *Q=1* followed by a second topological phase transition where the topological number changes from *Q=1* to *Q=0*. However, beyond a critical value of the interlayer interaction (see Figure 3) the bilayer germanene behaves as a single layer and the topological number of the twisted bilayer changes from *Q=2* to *Q=1*. In this case, there is only one electric field-induced topological phase transition from a topological insulator (*Q=1*) to a trivial band insulator (*Q=0*). As we have shown previously, there is compelling experimental evidence for the first topological phase transition in germanene [21]. The second topological phase transition in twisted germanene occurs, however, at a substantially higher critical electric field. For example, for an interlayer coupling parameter of 0.3 eV, this critical electric field is estimated to be about 2.5 V/nm. Unfortunately, this high electric field often results in damaging of the germanene layers. For smaller fields (about 1.95 V/nm) we observe that the twisted bilayer germanene has well-defined edge states, see Figure 2(c), indicating that the material is topologically non-trivial.

**Conclusions**

In conclusion, we have studied the electronic band structure of 21.8° twisted bilayer germanene. The crossing of the Dirac cones of the bottom and top germanene layers results in two Van Hove singularities with a gap of ~1.84 eV. Using this Van Hove gap we extract a Fermi velocity of $(8 \pm 1) \times 10^5 \ m/s$. A spatially resolved study reveals that the AA stacked sites are gapless, whereas the AB/BA stacked sites show a gap. The opening of the gap in AB and BA stacked sites is caused by an electric field in the tunnel junction, which arises due to a difference in work function between the scanning tunneling microscopy tip and substrate. Our tight-binding calculations show that twisted bilayer germanene should undergo two topological phase transitions from a topological number Q=2 to Q=1 and from Q=1 to Q=0, as a function of the applied electric field and interlayer coupling. Our *dI(V)/dV* spectra reveal a well-defined edge state (for an electric field of 1.95 V/nm), indicating that twisted bilayer germanene is indeed topologically non-trivial.

**Sample preparation and experimental methods**

The experiments are performed in an ultra-high vacuum system with a base pressure of $1 \times 10^{-11}$ mbar. The system is equipped with an Omicron low-temperature scanning tunneling microscope (STM). We have used Pt/Ir scanning tunneling microscope tips. Current-voltage *I(V)* curves are obtained in the constant height mode, i.e. the tunnel current is measured while the sample bias is ramped, with the feedback loop of the STM disabled. The *I(V)* curves that we used for our analysis are obtained by averaging over many *I(V)* traces that are collected in the grid scan mode. The differential conductivity (*dI(V)/dV*) is obtained by taking the numerical derivative of the *I(V)* curve. The STM experiments are performed at 77 K. The germanene layers and twisted bilayers are grown on $Ge_2Pt$ clusters. As a substrate, we used lightly doped *n*-type Ge(110). The Ge (110) samples are cut from nominally flat 10x10x0.4 mm, single-side-polished substrates. The Ge(110) samples are mounted on a Mo sample holder. Contact of the Ge(110) samples to any other metal during preparation and experiment is carefully avoided. After cleaning and mounting, the Ge(110) substrates are outgassed in ultra-high vacuum at 700 K for 24 hours. Subsequently, the Ge(110) substrates are cleaned by a cleaning method that has been successfully applied to the more abundant Ge(001) and Ge(111) substrates [56]. This method involves several cycles of Argon ion sputtering at 500-800 eV and annealing at 1100 (±25) K. After checking the cleanliness of the Ge(110) substrate with STM, a few monolayers of Pt are deposited onto the sample at room temperature. We used a home-built Pt evaporator, which consists of a W filament wrapped with a high-purity Pt (99.995%) wire. After Pt-deposition, the sample is annealed at 1100 (±25) K for one minute and subsequently slowly cooled down to room temperature. At temperatures exceeding the eutectic temperature (1043 K), eutectic $Ge_{0.78}Pt_{0.22}$ droplets form on the Ge(110) surface. Below the eutectic temperature, the $Ge_{0.78}Pt_{0.22}$ droplets undergo spinodal decomposition and separate into the two phases that are adjacent to the eutectic phase in the Ge/Pt phase diagram, i.e. pure Ge and $Ge_2Pt$. As the eutectic phase contains more Ge than the $Ge_2Pt$ alloy, the excess Ge is expelled to the surface of the solidified droplets and forms germanene. For more detailed information regarding the formation of germanene via the spinodal decomposition of eutectic $Ge_{0.78}Pt_{0.22}$ droplets we refer to refs. [20, 20-32] as well as the supplementary information.

**Tight-binding calculations**

The simplest model for a honeycomb system with spin-orbit interaction is the Kane-Mele model [57,58]. Germanene has a buckled honeycomb lattice and is well-described by the Kane-Mele model [59,60]. The tight-binding model describing bilayer germanene under perpendicular electric field $E_z$ is given by,

$$\hat{H} = -t \sum_{<i,j>su} c^\dagger_{isu} c_{jsu} - t_\perp \sum_{<i,j>su} c^\dagger_{isu} c_{js\bar{u}} + i\frac{\lambda_{SO}}{3\sqrt{3}} \sum_{\ll i,j \gg s} s\nu_{ij} c^\dagger_{isu} c_{jsu} - l \sum_{is} \mu_i E_z c^\dagger_{isu} c_{isu}$$

where $\langle\langle i, j \rangle\rangle$ runs over all the next-nearest neighbor hopping sites, $s$ refers to the spin index and $u$ to the layer index ($u$ and $\bar{u}$). $t$ and $t_\perp$ are the intralayer and interlayer hopping parameters, $\lambda_{SO}$ is the spin-orbit coupling, $l$ the buckling and $E_z$ the electric field. In the supplementary information we have taken $t = 1.04\ eV$ and $\lambda_{SO} = 43\ meV$ from refs. [59,60], whereas in the main text we have used $t = 0.92\ eV$ and $\lambda_{SO} = 70\ meV$ as they fit the experimental observations better. The fourth term represents the effective spin-orbit coupling, where $\nu_{ij} = 1$ refers to hopping in the clockwise direction and $\nu_{ij} = -1$ to hopping in the anti-clockwise direction with respect to the positive z-axis. The last term represents the staggered sublattice potential with $\mu_i = 1$ for an A site and $\mu_i = -1$ for a B site.

**Acknowledgements**

PB acknowledges NWO Veni for financial support. HJWZ acknowledges the research program "Materials for the Quantum Age" (QuMat) for financial support. This program (registration number 024.005.006) is part of the Gravitation program financed by the Dutch Ministry of Education, Culture and Science (OCW). CC, DJK, and HJWZ acknowledges NWO (Grants 16PR3237 and OCENW.M20.232) for financial support. ME acknowledges CREST, JST (Grant JPMJCR20T2) and Grants-in-Aid for Scientific Research from MEXT KAKENHI (Grant No. 23H00171) for financial support.

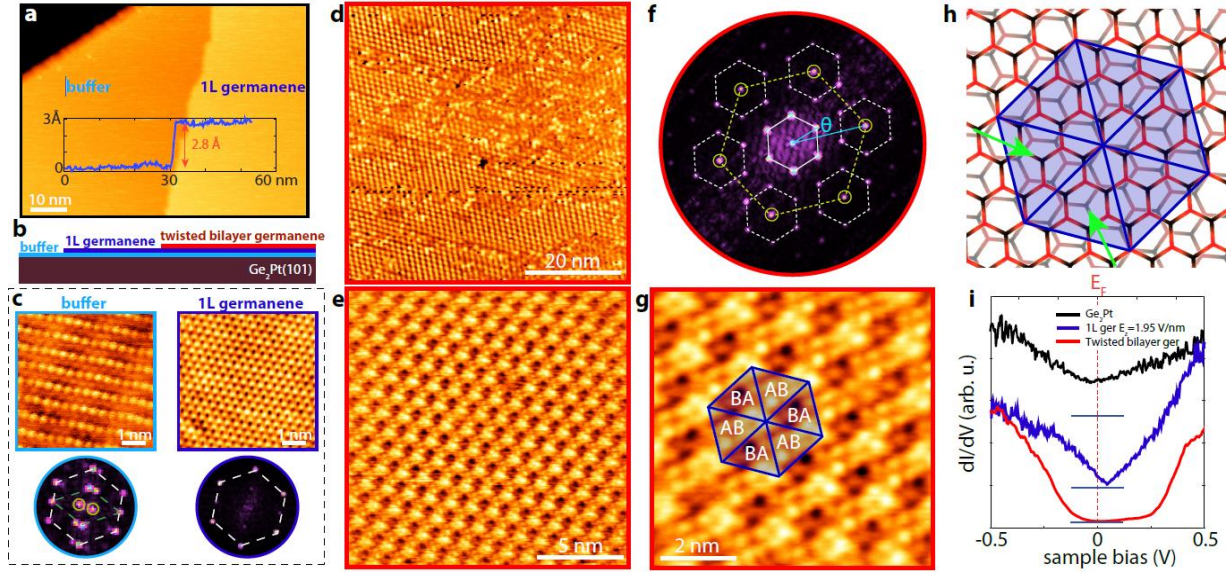

*Figure 1* (a) Large scale scanning tunneling microscopy image of few-layer germanene on Ge$_2$Pt. Inset: line profile across the step edge with a monoatomic germanene height of 0.28 nm. Sample bias -1.5 V and tunnel current 0.2 nA, (b) Cartoon of the different germanene layers (buffer layer, 1L germanene and twisted bilayer germanene) on Ge$_2$Pt(101) (c) Small-scale STM images and the corresponding FFT images of a buffer layer germanene (left) and a monolayer (1L) germanene on buffer/Ge$_2$Pt (right). The top left panel shows germanene (honeycomb), Ge$_2$Pt (rhombic) and moiré lattices (one-dimensional). The top right panel shows the low-buckled honeycomb lattice of germanene. Sample bias -1.3 V and tunnel current 0.3 nA. (d) A large-scale image on a few-layer germanene/Ge$_2$Pt cluster, showing a larger hexagonal periodicity. Sample bias -1.5 V and tunnel current 0.2 nA. (e) A zoom-in on one of these regions revealing a complex structure corresponding to a 21.8° twisted bilayer germanene. Sample bias -1.3 V and tunnel current 0.3 nA. (f) Fast Fourier transform of the image shown in panel (e). θ is the angle between the honeycomb lattice of the top layer germanene and the moiré lattice. The yellow and white hexagons represent the reciprocal lattice of the germanene honeycomb lattice and the moiré lattice, respectively. The dotted circles are replicas of the moiré unit cell at the first order spots of reciprocal lattice of germanene. (g) Atomic resolution STM image (high-pass filtered) of a 21.8° twisted bilayer germanene, showing the AA, AB and BA sites. Sample bias -1.3 V and tunnel current 0.3 nA. (h) Structural model of 21.8° twisted bilayer buckled germanene. The green arrows refer to the locations where only one atom is observed in either the top layer or the bottom layer. (i) dI(V)/dV spectra recorded on Ge$_2$Pt, 1L germanene and twisted bilayer germanene for a tip induced electric field of approximately 2 V/nm. Setpoint sample bias -0.5 V and setpoint tunnel current 1 nA.

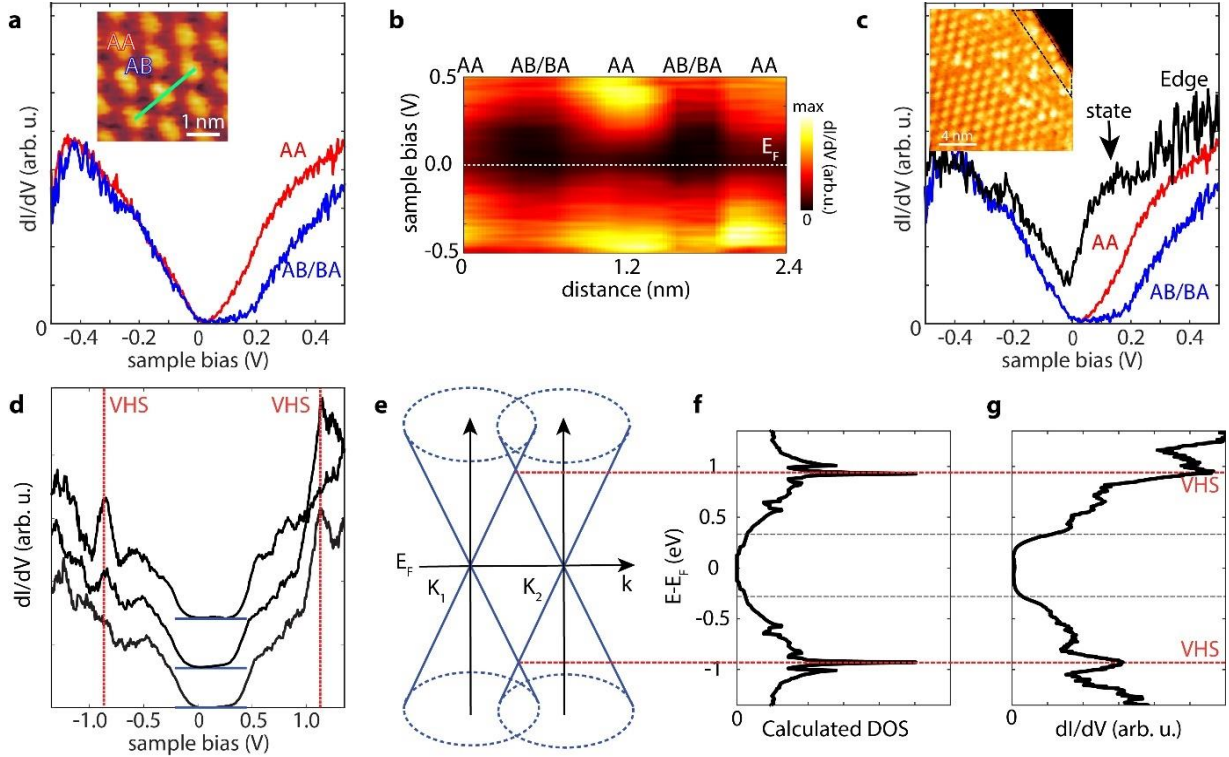

*Figure 2* (a) Differential conductivity (dI(V)/dV) of an AB/BA stacked area showing a band gap with a width of ∼ 0.2 eV (blue line) and a AA stacked area without a band gap (red line). Inset: Small scale STM image of 21.8° twisted bilayer germanene. The bright spots are AA stacked regions. Setpoint sample bias -0.5 V and setpoint tunnel current 1 nA. (b) dI(V)/dV recorded along the green line shown in panel (a), showing that the size of the band gap is modulated by the moiré pattern. (c) dI(V)/dV spectra recorded at the edge (black) and compared to the AA (red) and AB/BA (blue) sites. Inset: topography showing the edge of the twisted bilayer, outlined by the dashed black box. Setpoint sample bias -0.5 V and setpoint tunnel current 1 nA, (d) dI(V)/dV spectra of 21.8° twisted bilayer germanene versus sample bias recorded at various locations. The two Van Hove singularities (VHS), which arise due to the crossing of the Dirac cones of the two germanene layers, are located at -0.8 eV and 1.04 eV, in all curves. Setpoint sample bias -1.5 V and setpoint tunnel current 1 nA, (e) Schematic diagram of the formation of Van Hove singularities in twisted bilayer germanene. Calculated (f) and measured (g) density of states versus energy of 21.8° twisted bilayer germanene. We have chosen an intralayer hopping parameter of 0.92 eV, an interlayer hopping parameter of 0.3 eV, an interlayer bias of 0.5 V and a spin-orbit coupling of 70 meV for the tight-binding calculations. The calculated density of states is shifted such that the Fermi level coincides with the center of the bandgap in the dI/dV spectrum. The red dotted lines mark the locations of the Van Hove singularities.

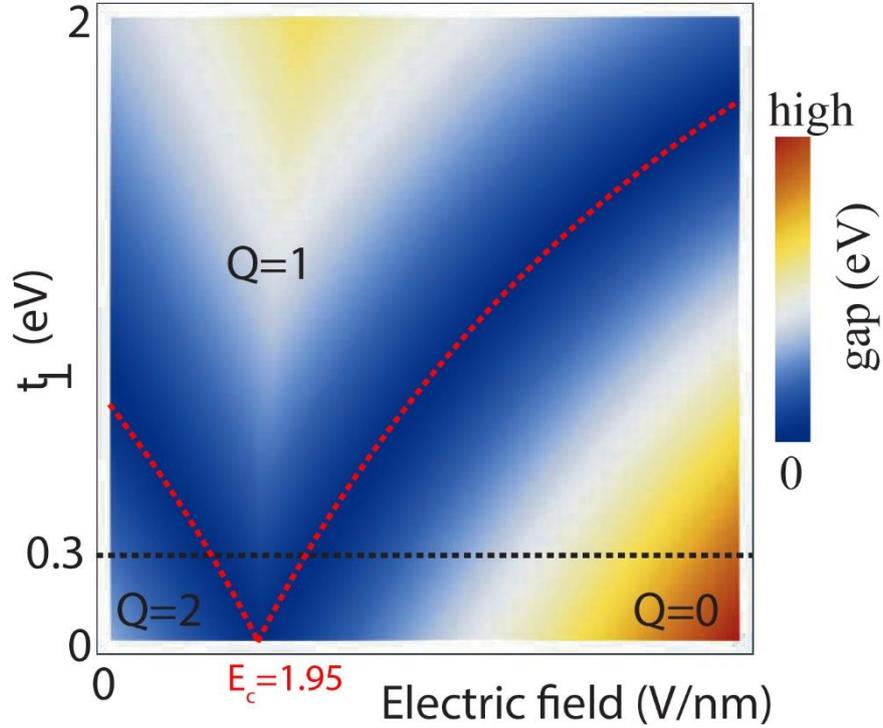

*Figure 3* Topological phase diagram of twisted bilayer germanene, showing the band gap as a function of the applied electric field (x-axis) and interlayer hopping parameter (y-axis). $E_c$ stands for the critical field of germanene in the case of a vanishing interlayer interaction. The red dotted lines mark the topological phase transitions from Q=2 to Q=1 and Q=1 to Q=0, respectively. The dotted black line refers to an interlayer hopping of 0.3 eV. The numerical data should be considered as a guideline rather than accurate values. Moreover, our estimate of the electric field contains experimental offsets that are difficult to disentangle. For detailed insights, refer to references [21, 55].